\documentstyle[psfig,graphicx,rotate]{article}
\begin{document}
\vspace*{-2.5cm}
%to be submitted to JAA

\begin{center}
{\bf \Large   Unravelling the nature of HD 81032 - a new RS CVn Binary}
\end{center}

\begin{center}
    J. C. Pandey$^1$\footnote{E-mails:jeewan@aries.ernet.in; singh@tifr.res.in;drake@olegacy.gsfc.nasa.gov; sagar@aries.ernet.in}, K. P. Singh$^{2*}$, S. A. Drake$^{3*}$, and R.
    Sagar$^{1*}$
\end{center}

   $^1$ Aryabhatta Research Institute of Observational Sciences, Naini Tal - 263 129, India
  
   $^2$ Tata Institute of Fundamental Research, Mumbai - 400 005, India

   $^3$ USRA \& Code 662, NASA/Goddard Space Flight Center, Greenbelt, MD 20771, USA

\begin{abstract}
BVR photometric and quasi-simultaneous optical spectroscopic 
observations of the star HD 81032 have been carried out during the years 
2000 - 2004. A photometric 
period of $18.802 \pm 0.07$ d has been detected for this star.
A large group of spots with a migration period of $7.43 \pm 0.07$ years is inferred 
from the first three years of the data.
H$\alpha$ and Ca II H and K
emissions from the star indicate high chromospheric activity.
The available photometry in the BVRIJHK bands is consistent with spectral
type of K0 IV previously found for this star.
We have also examined the spectral energy distribution of HD 81032 for 
the presence of an infrared colour excess using the 2MASS JHK and IRAS 
photometry, but found no significant excess in any band above
the normal values expected for a star with this spectral type.
We have also analyzed the X-ray emission properties of this star using
data obtained by the ROSAT X-ray observatory during its All-Sky Survey
phase. An X-ray flare of about 12 hours duration was detected during the two
days of X-ray coverage obtained for this star. Its X-ray spectrum, while
only containing 345 counts, is inconsistent with a single-temperature
component solar-abundance coronal plasma model, but implies either the
presence of two or more plasma components, non-solar abundances,
or a combination of both of these properties. All of the above properties
of HD 81032 suggest that it is a newly identified, evolved
RS CVn binary. 

\end{abstract}


\baselineskip 16pt
%----------------------Introduction-------------------
\section{Introduction}
\label{int.sec}
The presence of strong X-ray and nonthermal radio emission in late-type stars 
is a well-known
indication of enhanced coronal activity (Drake et al. 1992; G\"{u}del 2002). 
X-ray  emission from coronae ($T_e \sim 10^{6-7}$K) and
chromospheric emission ($T_e \sim 10^4$K) are also closely correlated, and thus stars with
intense coronae will have strong chromospheric emission, as evidenced by their UV and
Ca II H and K line emission (e.g. Ayres et al. 1995).
 While many active stars
have been identified as such through their X-ray and radio
emission, it is only through detailed optical photometric and
spectroscopic studies that their activity can be classified into known types.

RS CVn binaries are perhaps the most common type of active star, at least 
for stars having spectral types in the range from late F through K. 
In an RS CVn binary, the two 
stars  are usually tidally locked so that the rotational period of 
each star is approximately the same as the orbital period; however,
the two stars are not undergoing mass transfer (i.e. they form a detached
system). One star is generally a spectral type F to G
dwarf or a subgiant that is $\sim$ 1000 K hotter than its companion, which is usually a G to K type giant or 
sub-giant.  In the more recent definition of the RS CVn class, there is no 
restriction on the 
spectral type of the secondary star or on the orbital period, except for the 
evolutionary constraint (Fekel et al. 1986). 
RS CVns have been further subdivided into three groups according 
to their orbital period: the short period ($P \leq 1$ d),
classical ($1 \rm{d} \leq P \leq 14$ d), and long period ($P\geq14$ d) groups. 
In the soft (0.1 - 2.0 keV) X-ray band, the luminosities of RS CVn binaries
typically lie in the range of $10^{29} - 10^{32}$ erg s$^{-1}$ (Drake et al. 
1989), compared to $10^{25.5} - 10^{29.5}$ erg s$^{-1}$ 
for normal late-type stars (Schmitt \& Liefke 2004).
VLA and ATCA observations at cm wavelengths have shown that the radio 
luminosities of RS CVn systems are in the range from $10^{14.5}$ to 
$10^{17.5}$  erg s$^{-1}$ Hz$^{-1}$ (Drake \& Linsky 1986, Drake et al.1989), which is
also enhanced compared to that found for other late-type stars, typically
$\leq 10^{15}$ erg s$^{-1}$ Hz$^{-1}$ (G\"{u}del 2002).

The K0 IV star HD 81032 (Houk \& Smith-Moore 1988) was first noticed as an 
X-ray active star when it was identified as the likely optical counterpart 
of a soft X-ray (0.2-4.0 keV) source in the Einstein Slew Survey, 
1ES 0920-13.6, with  an X-ray flux of $2.1\pm0.9 \times 10^{-11}$ erg 
cm$^{-2}$ s$^{-1}$ (Elvis et al. 1992, Schachter et al. 1996). An X-ray 
source at this position was later detected  in the ROSAT All-Sky-Survey (RASS) 
at a weaker (but statistically more significant) level of $3.9\pm0.4 
\times 10^{-12} $ erg cm$^{-2}$ s$^{-1}$,
named 1RXS J092253.7-134919 in the RASS Bright Source Catalogue 
(Voges et al. 1999). This star was detected  on February 12, 1993 as a radio source,
with a 3.6 cm flux of $0.68 \pm 0.05$ mJy (Drake, S. A., private communication).
Given its spectral type and luminosity class, and its V-magnitude of
8.91 (Wright et al. 2003), the spectroscopic distance of HD 81032 is 
$140 \pm 45$ pc, implying an X-ray luminosity of $9.2 \pm 4.2 \times 10^{30}$ 
erg s$^{-1}$ (based on the RASS count rate and an assumed conversion factor
of $6 \times 10^{-12}$ erg cm$^{-2}$ PSPC cts$^{-1}$: a self-consistent
X-ray luminosity based on the RASS data is calculated in \S 8), 
and  a radio luminosity of $1.6 \pm 0.7 \times 10^{16}$ erg s$^{-1}$ Hz$^{-1}$,
where the distance uncertainty is the dominant contributor to the large errors
in the luminosities. These values clearly show that the star HD 81032 
has an active corona.

In this paper, we present extensive optical photometric and spectroscopic 
observations, as well as an analysis of archival X-ray data,  of the star 
HD 81032. This is the first detailed optical photometric and spectroscopic 
study of this star, although a brief discussion of some of the optical 
photometry of this star was given in Pandey et al. (2002). Based on our 
long-term optical study, the star HD 81032 is shown to be a new member of 
the (long-period) RS CVn class.

The organization of this paper is as follows. The observations and data 
reduction are presented in \S ~\ref{obsdat.sec}. In \S ~\ref{ligper.sec}, 
we analyse the data for the periodicity. In \S ~\ref{phopha.sec}, we 
discuss the light curve and phase of minimum light. Chromospheric emission 
features are described in \S \ref{halcai.sec}. In \S \ref{physical.sec} and 
\S \ref{sed.sec}, the spectral type and IR excess of HD 81032 are discussed, 
while the X-ray spectra are discussed in \S \ref{xspec.sec}. Finally, 
\S ~\ref{con.sec} summarizes the main results of this paper.

\section{Observations and data reduction}
\label{obsdat.sec}

\subsection{Optical Photometry}
HD 81032 was  observed in the Johnson B, V and Cousins R filter for 122 
nights during four observing runs -
year 2000-2001 (19 nights), year 2001-2002 (54 nights), year 2002-2003 
(24 nights) and year 2003-2004 (25 nights) - 
at the Aryabhatta Research Institute of Observational Sciences (ARIES).
The observations were made with the 104-cm Sampurnanand telescope and using
a $2k\times2k$  CCD camera in years 2000 - 2003, and a $1k\times1k$ CCD 
camera during the years 2003 - 2004.
A few  CCD frames were taken in the B, V and R filters on every night, with 
exposure times ranging from 2
to 60 secs, depending upon the seeing conditions and the filter used.
Several bias and twilight flat frames were also taken during each observing run.
Bias subtraction, flat fielding and aperture photometry were performed using
IRAF\footnote{IRAF is distributed by National Optical Astronomy 
Observatories, USA}.  For the star HD 81032 the comparison and check stars 
were TYC 5471 1345 1 and USNO-A2.0 0750-06845737, respectively.
Differential photometry, in the sense of variable minus the comparison 
star, was done, since  program, comparison and check stars 
were all in the same CCD frame.
UBVRI observations of HD 81032 along with some of the Landolt (1992)
standard region SA 98 were obtained on 22 February 2004 for photometric
calibration. The average magnitudes of HD 81032 during the observing years 
2003 - 2004 in 
the U, B, V, R and I filters
were $10.242 \pm 0.007, 9.635 \pm 0.005, 8.614 \pm 0.005, 8.401 \pm 0.003,$
and $7.736 \pm 0.003$,  respectively.

\subsection{Optical Spectroscopy}
\label{spec.sec}
Spectroscopic observations were carried out during 2003 January 20 to 24 
at the Vainu Bappu Observatory, Kavalur with the OMR spectrograph fed by
the 234-cm Vainu Bappu Telescope. 
The data were acquired with a $1024\times1024$ CCD camera
of $ 24\times 24 \mu m$ square pixel size covering a range of $1200 \AA$ and
having a dispersion of $1.25 \AA/pixel$ . 
The star was observed in the wavelength ranges of
3500 - 4700 \AA ~and 5700 - 6900 \AA.
Four spectra of HD 81032 in the H$\alpha$ region
and three in the CaII H and K region were obtained.
a signal-to-noise of ratio between 20 to 40 was achieved in these spectra. 
HD 71952, a K0IV type star, was also observed as a reference star for
HD 81032.

The spectra were extracted using the standard reduction procedures in the
IRAF packages (bias subtraction, flat fielding, extraction of the spectrum
and wavelength  calibration using arc lamps). The spectral resolution
was determined by using emission lines of arc lamps taken on the same nights.
A spectral resolution ($\delta \lambda$) of $2.7 \AA$ at $6300 \AA$ and
$3.7 \AA$ at $4000 \AA$ was achieved.
All the spectra were normalized to the continuum and equivalent widths
for the emission lines were computed using the IRAF task {\it splot}.
The errors in the measurement of the equivalent widths of the emission lines
were determined by measuring the equivalent widths of some moderate absorption
lines for each spectra.
We computed the standard deviation for the equivalent
width of each feature, and finally determined the mean standard deviation 
to be $ 0.02 \AA$.

\subsection{X-ray Data}
The star HD 81032 was observed and detected by the ROSAT PSPC
detector during  the ROSAT All-Sky-Survey (RASS) phase over a 2-day period
from 1990 November 10 to 12.
The exposure time was 501 s, and was accumulated in 26 separate short scans
of this region of sky. The PSPC had an energy 
range from 0.1 - 2.4 keV with a (low) spectral resolution ($\Delta E/E 
\approx 0.42$ at 1 keV). A full description of the X-ray telescope and 
detectors can be
found in Tr\"{u}mper (1983) and in Pfeffermann et al.(1987).
The ROSAT X-ray data for HD 81032 were obtained from the public archives,
the relevant RASS dataset being rs932025n00. 
%The off-set of HD 81032 from the field center is 187.8'.
Source spectra for HD 81032 were accumulated from on-source counts obtained 
from a circular region on the sky centered on the X-ray peak and having 
a radius of 3.85 arcmin. The background was
accumulated from several neighboring regions at nearly the same offset
from the source.

\section{Photometric light curves and period analysis}
\label{ligper.sec}
\begin{figure}
\centering
\vbox{
\includegraphics[height=4.6cm,width=9.5cm]{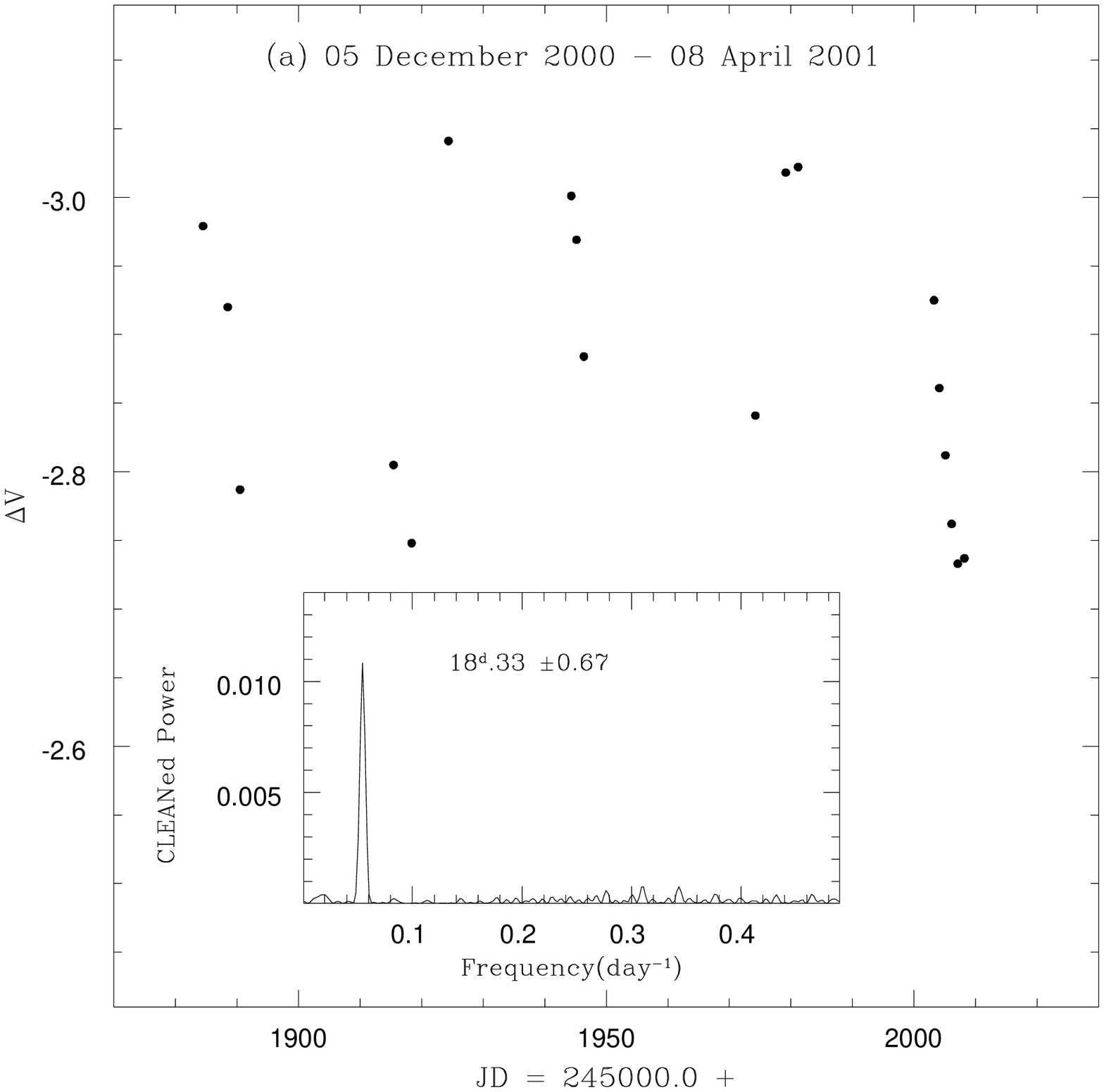}
\includegraphics[height=4.6cm,width=9.5cm]{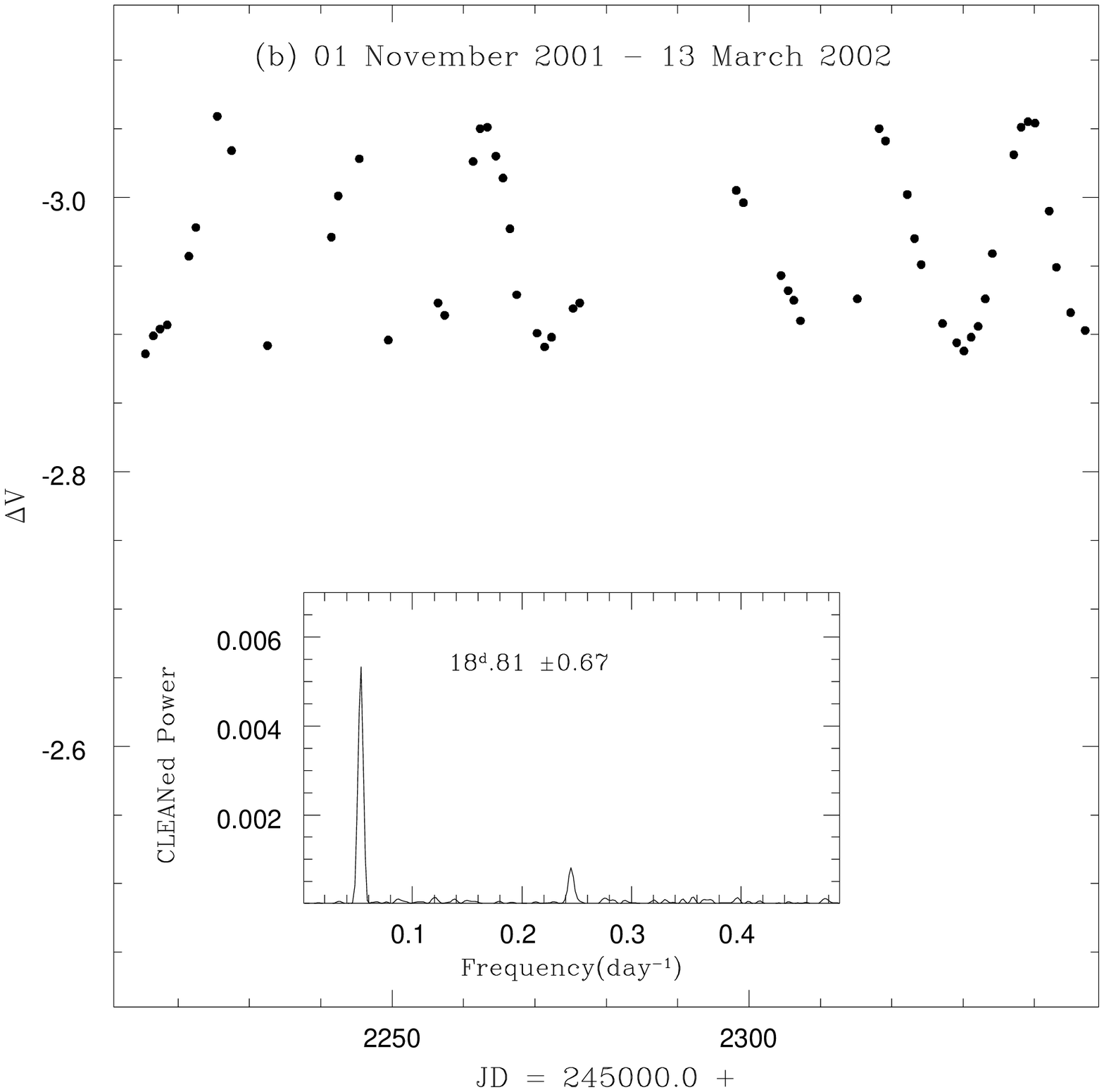}
\includegraphics[height=4.6cm,width=9.5cm]{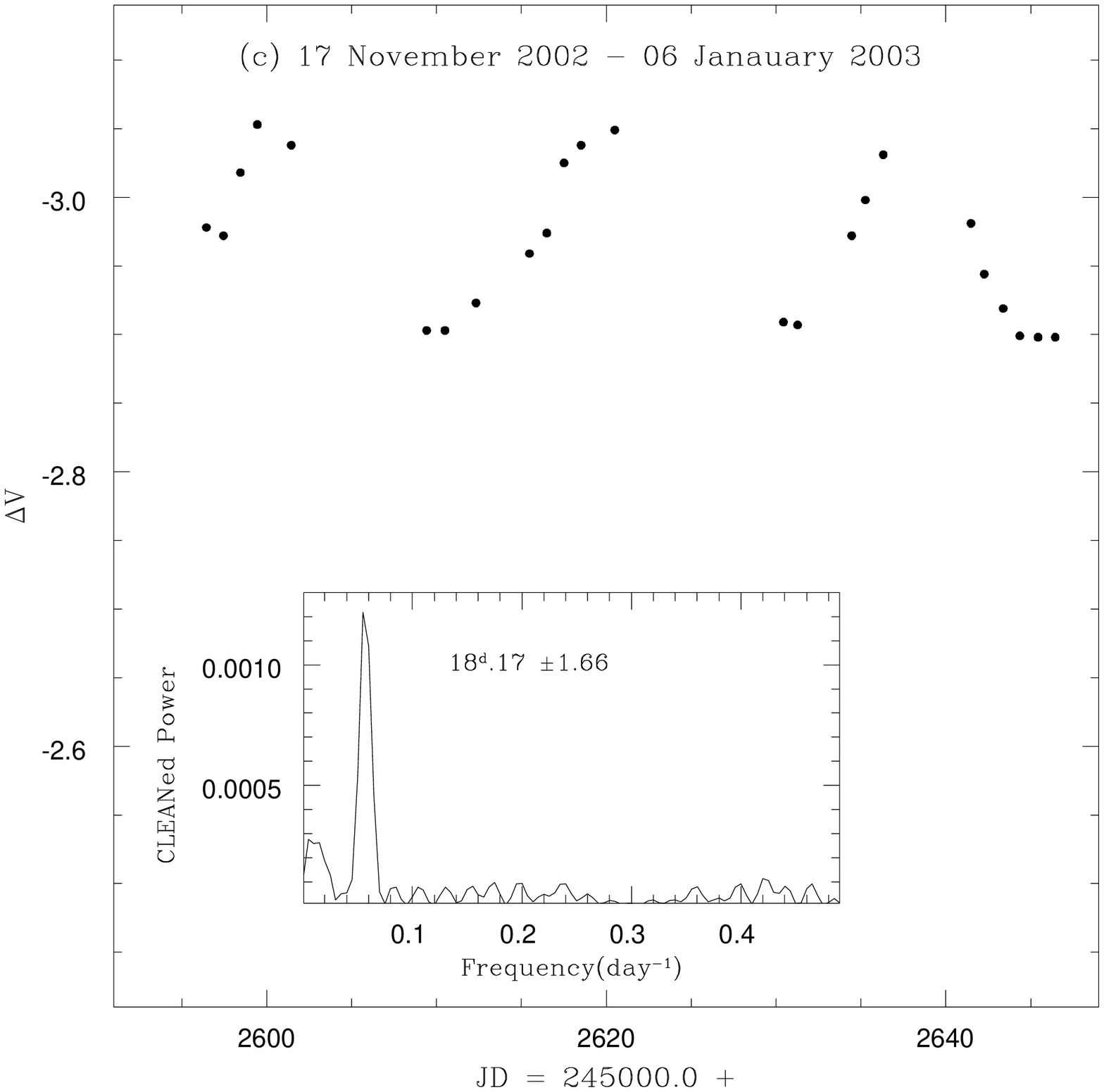}
\includegraphics[height=4.6cm,width=9.5cm]{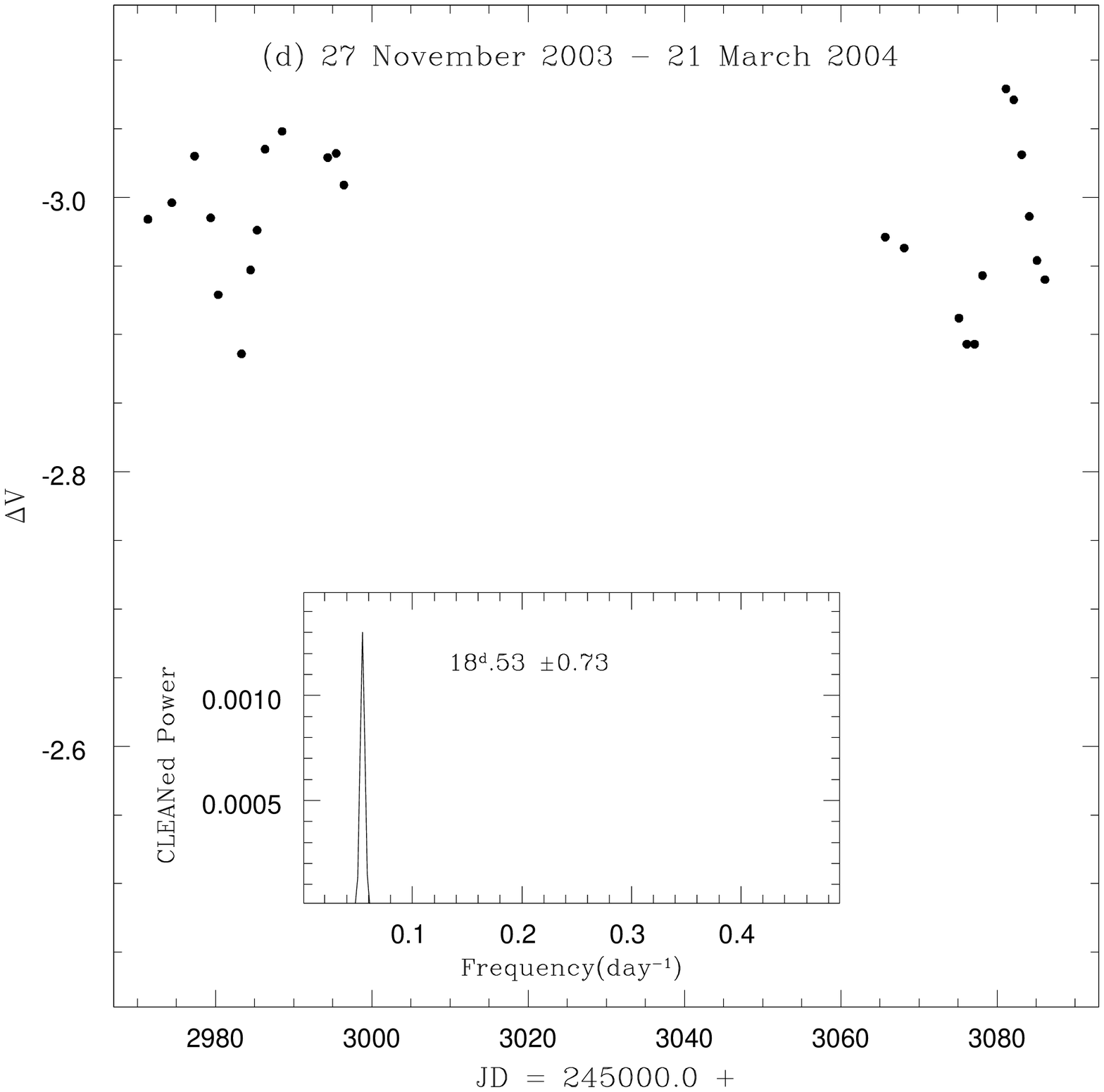}
}
\caption{The V band light curves and corresponding CLEANed power density spectra 
(insets). The epoch is given at the top of each panel, and the 
period is written at the top of each inset.}
\label{ligper.fig}
\end{figure}

\begin{figure}
\centering
\includegraphics[height=9cm,width=12.0cm]{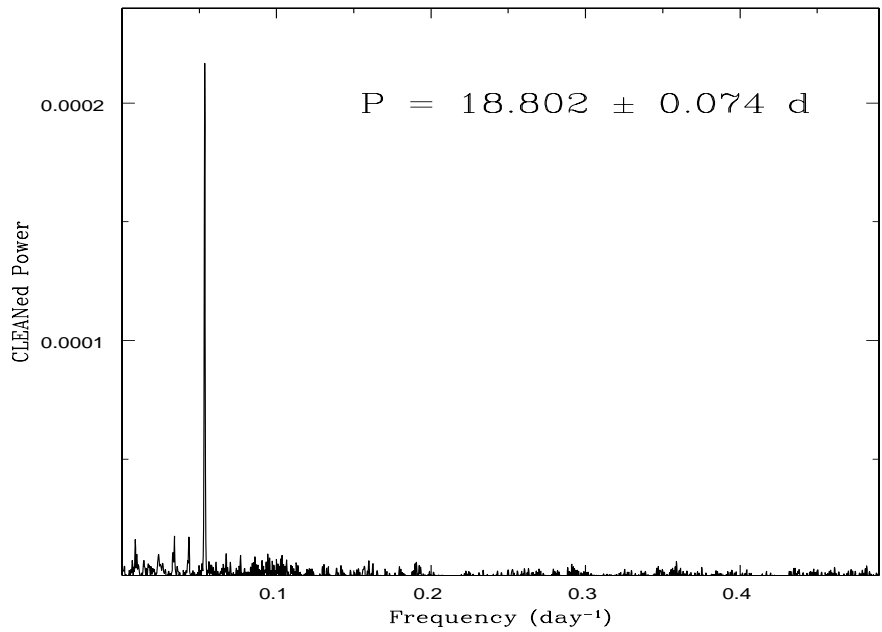}
\caption{ CLEANEd power spectra of whole data taken during 2000 - 2004 of HD 81032.}
\label{power.fig}
\end{figure}

Photometric light curves corresponding to the four observing runs were taken.
Figure \ref{ligper.fig} shows the  V band differential light curves of the star HD 81032 at different epochs. 
 We did not find
any significant variations in the comparison star (see below). 
%Lower panel of Fig. 4(a to e)
%shows the differential light curves of the comparison star in the sense of comparison minus check star.
The yearly mean of the standard deviation($\sigma$) between the different measures of 
comparison and check stars in the B, V and R filters was found to be 0.011, 0.01 and 0.01,
respectively. 
Each light curve shown in Fig. \ref{ligper.fig} was analysed for periodicity.
To find a period from unequally spaced data, we  used the 
CLEAN  algorithm (Roberts et al. 1987) in Starlink's PERIOD software. 
The power spectrum obtained using this method is shown in the inset of each panel of Fig. \ref{ligper.fig}
along with the period determined. The CLEANed power spectra presented were obtained 
after 100 iterations with a loop gain of 0.1. 
The period is found to be constant within error for each epoch.
To improve the  
period  determination of the star HD 81032, the entire data  from 2000 - 2004 were analysed
using the same algorithm.  Figure \ref{power.fig}  shows the CLEANed power spectrum from the entire dataset. 
The highest peak in the CLEANed power spectrum corresponds to a period of $18.802 \pm 0.074 \rm{d}$.
The 18.802 d period is much more plausible than the 57 d period reported  
 earlier (Pandey et al. 2002). The previous determination was 
mainly due to observational limitations as our early data were too 
sparse and highly uneven.
 Besides a subgiant like this 
one is unlikely to be synchronized in a binary of 57-days period.

\section{Photometric variation and phase of minima}
\label{phopha.sec}
\begin{figure}
\centering
\hbox{
\includegraphics[height=8.7cm,width=6cm]{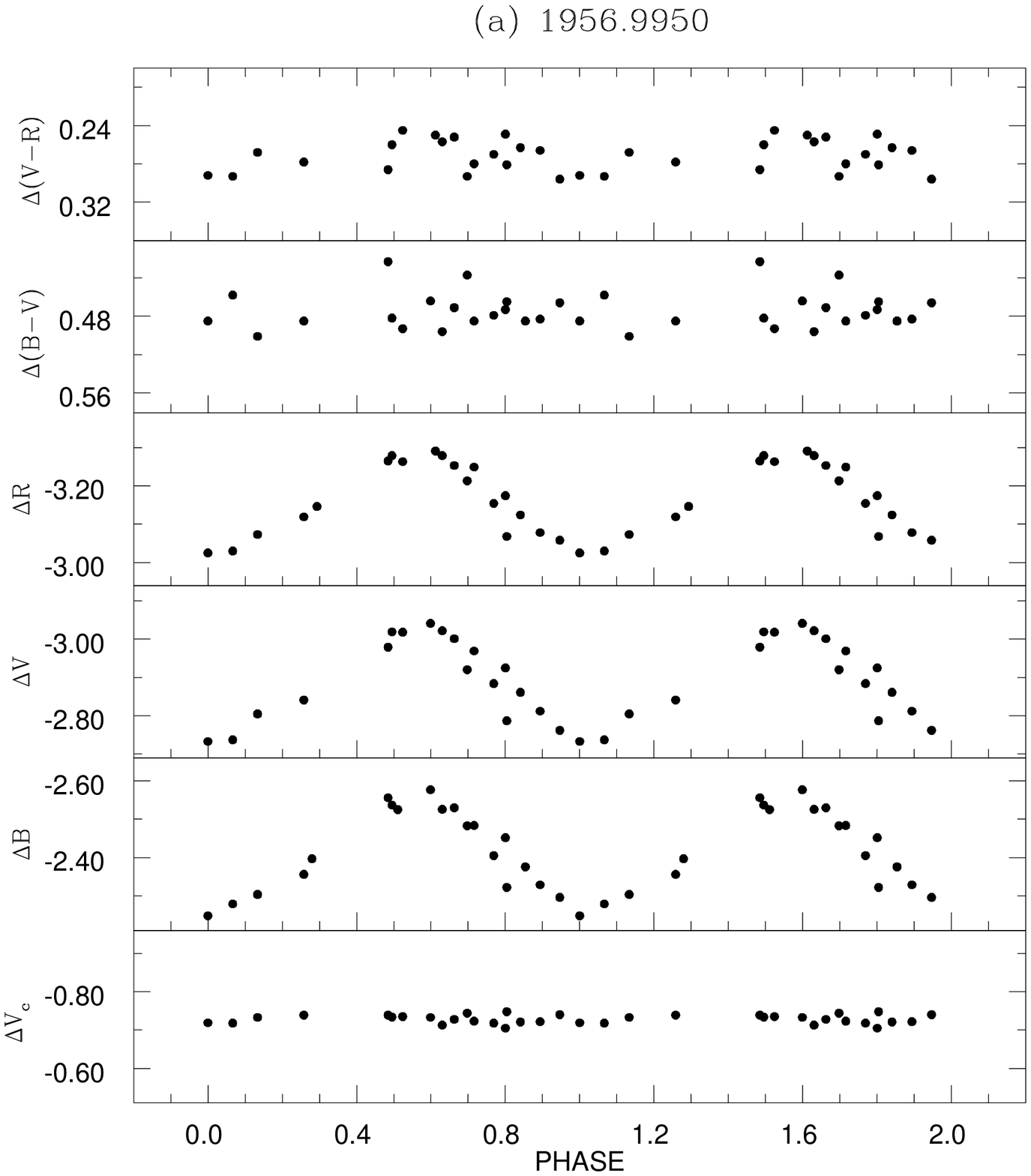}
\includegraphics[height=8.7cm,width=6cm]{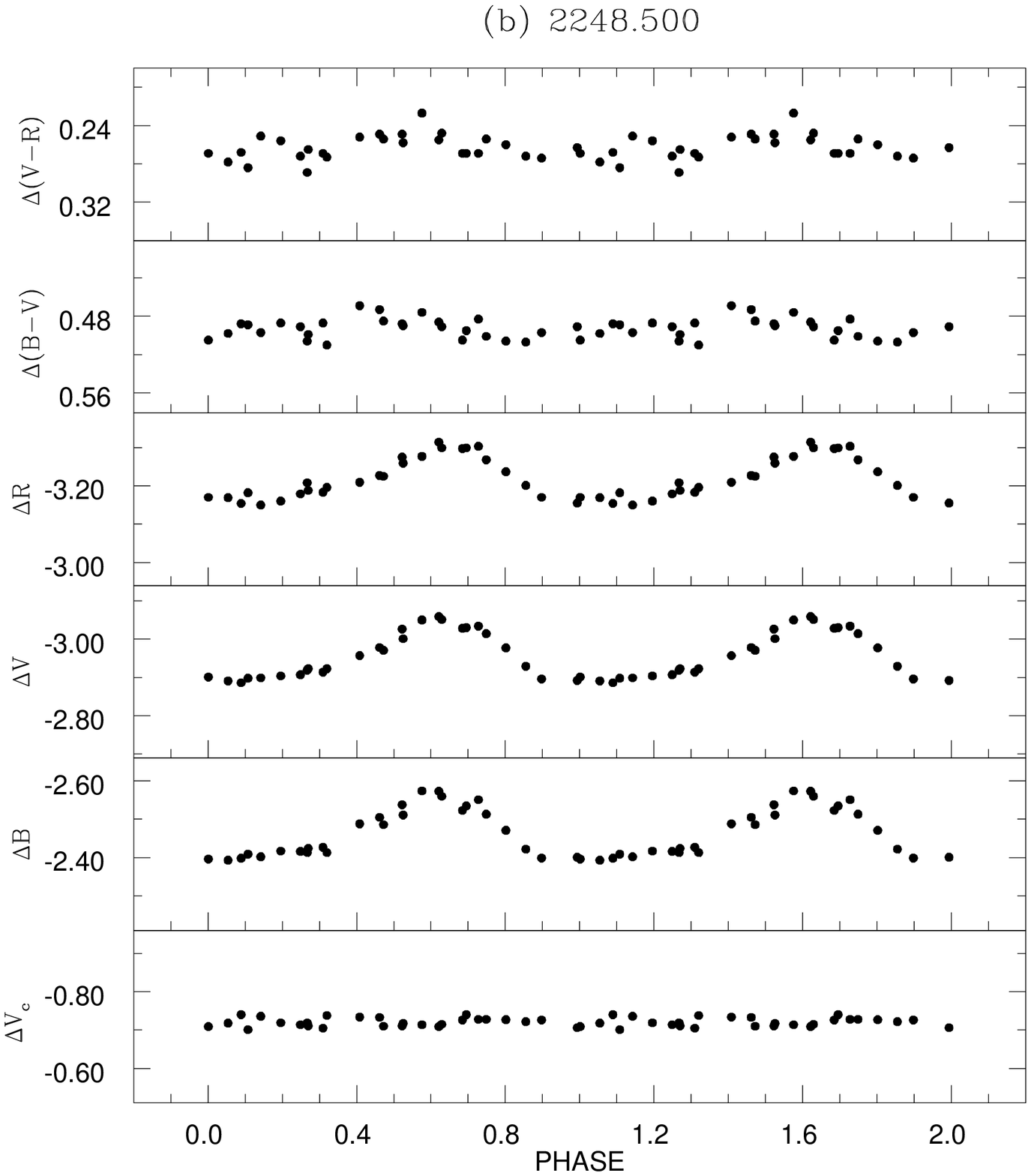}
}                                                

\vspace{0.5cm}

\hbox{                                           
\includegraphics[height=8.7cm,width=6cm]{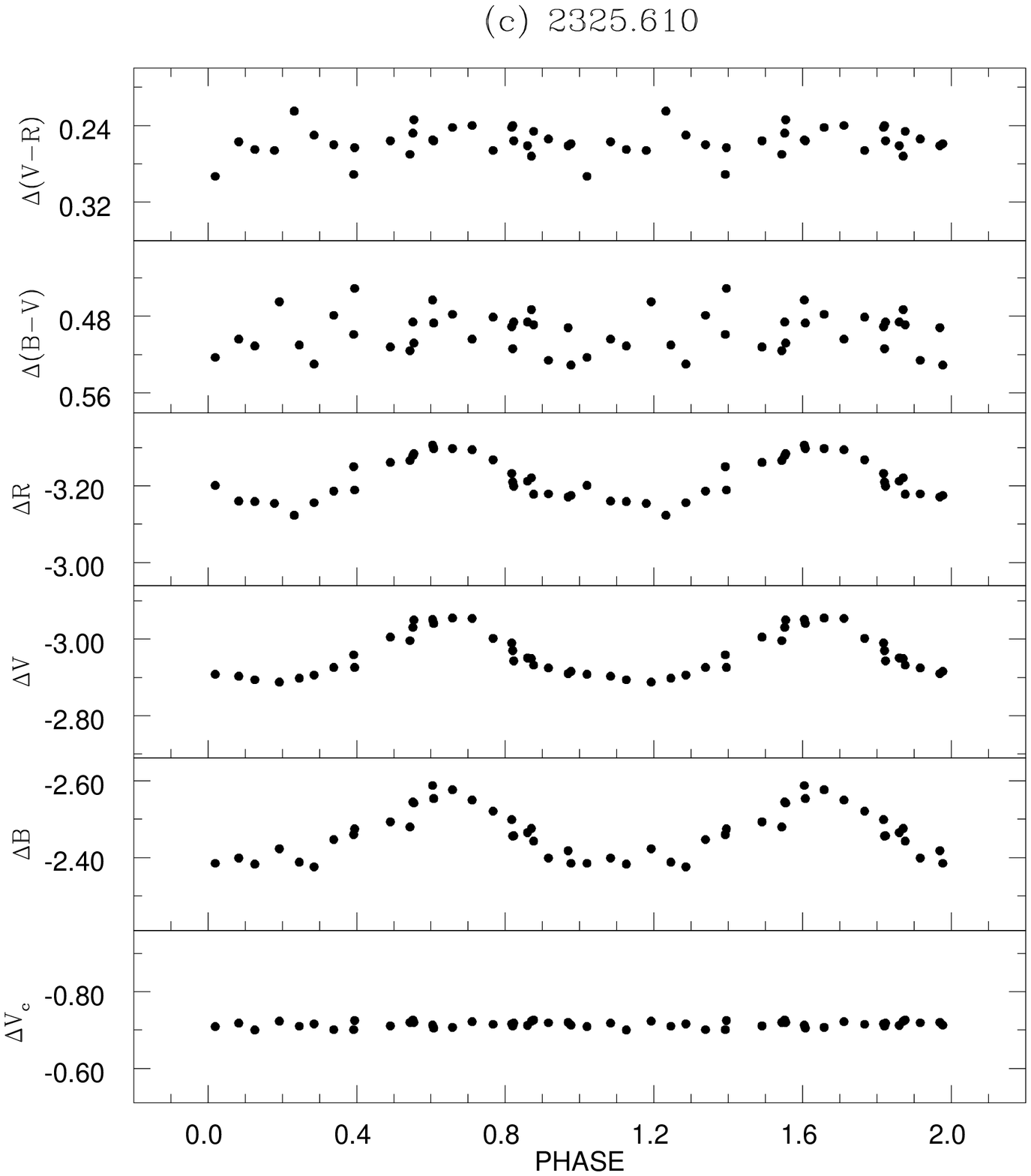}
\hspace{-0.5cm}
\includegraphics[height=8.7cm,width=6cm]{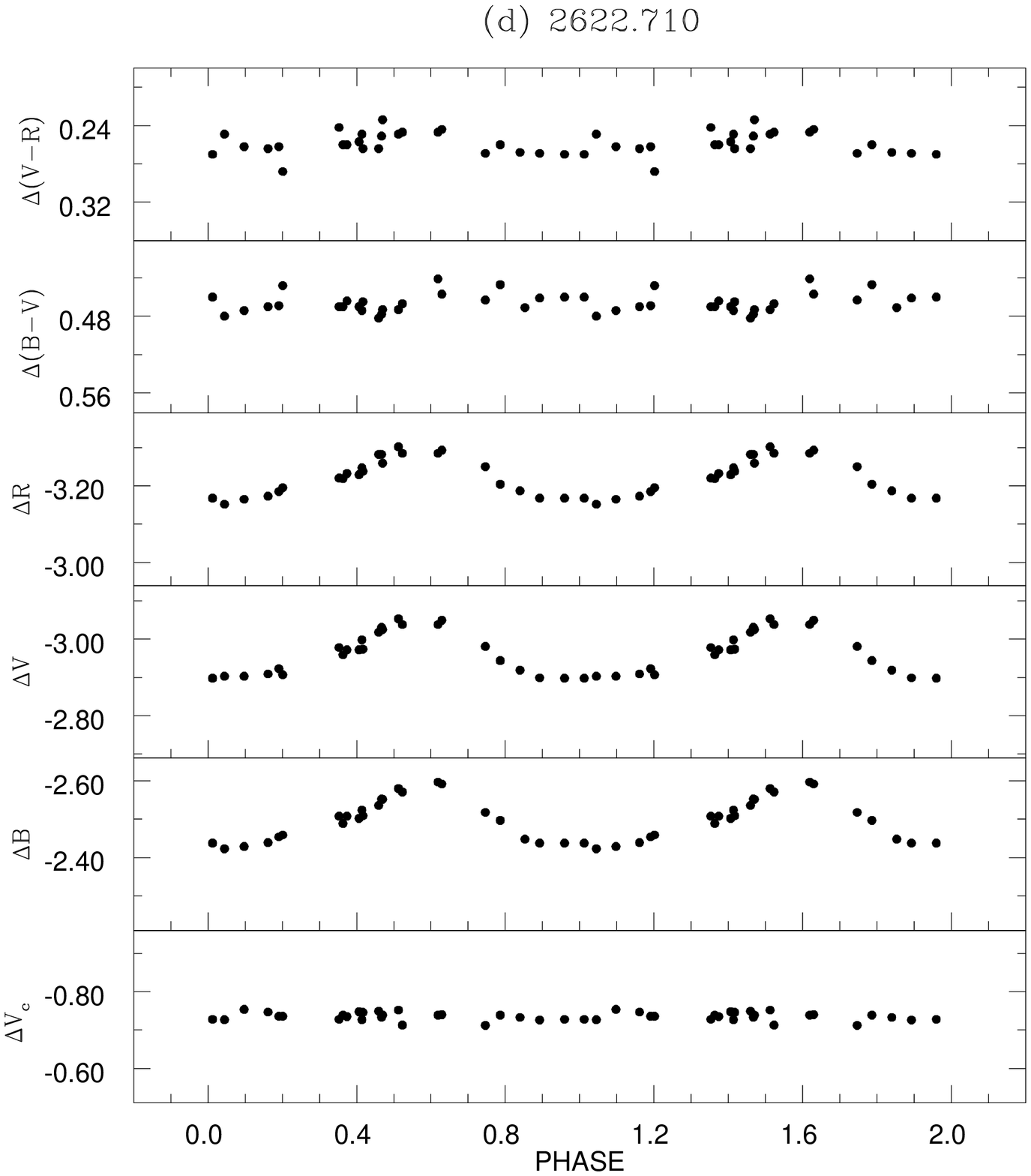}
\hspace{-0.5cm}
\includegraphics[height=8.7cm,width=6cm]{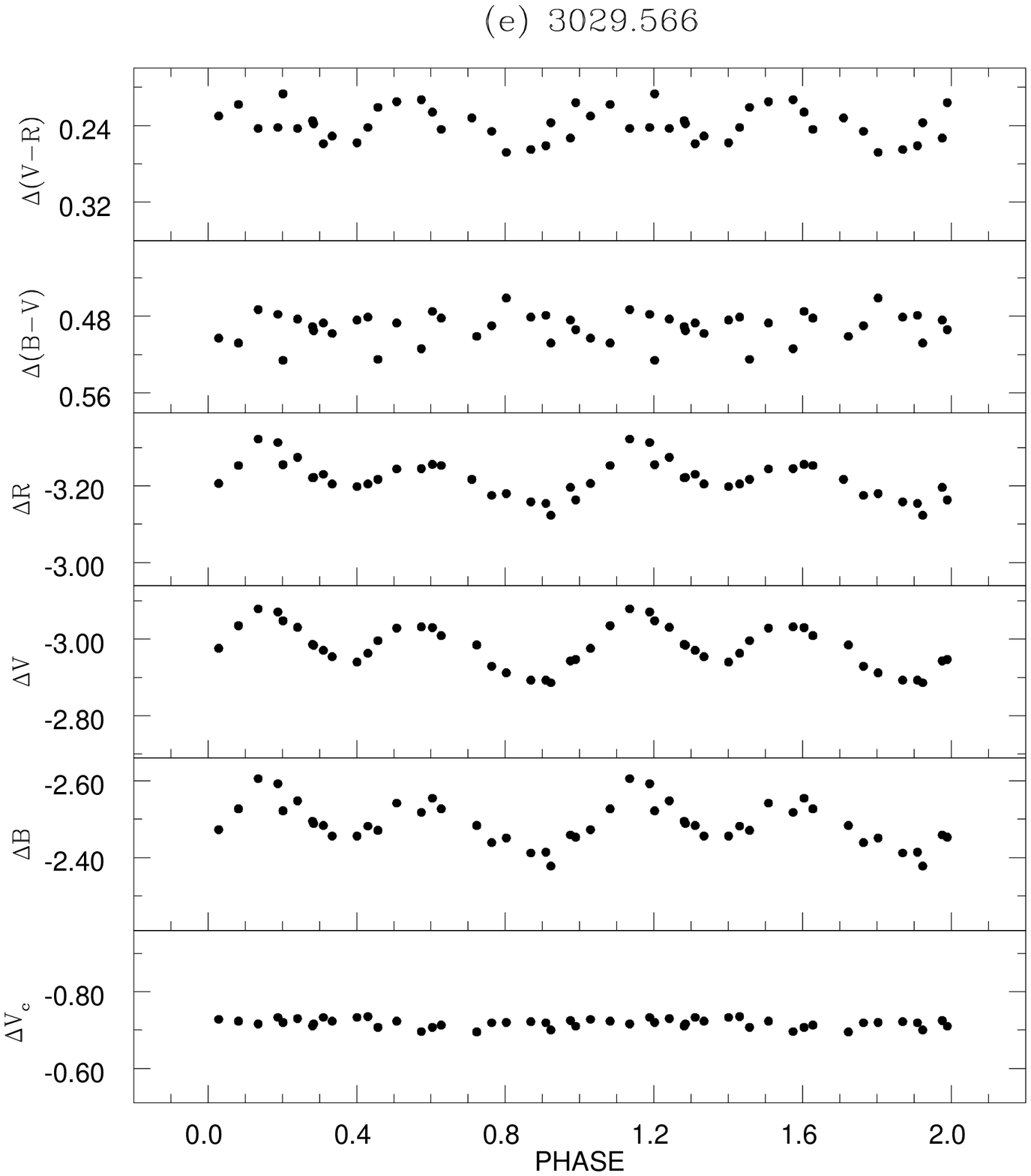}
}
\caption{Differential $V_{c},B, V, R$ light curve and $(B-V), (V-R)$ colour
curve of HD 81032. $\Delta V_{c}$ stands for the differential magnitude 
between comparison and check stars. The light curve is folded using the period 18.802 d. The epoch of
each light curve is mentioned at the top of each figure.} 
\label{fold.fig}
\end{figure}

The Julian days of the observations were converted to the phases using the ephemeris:

$ Phase(\theta) = JD 2452307.761 + 18^{d}.802  E$

where the initial epoch corresponds to the conjunction with the first minimum observed.
Figure \ref{fold.fig}  shows the differential B, V, R and V$_c$, light curves and, (B-V) and (V-R) 
colour curves  of the star HD 81032. Here V$_c$ stands for the differential V band light curve 
of the comparison and the check stars.
Each point in the light curves is  mean of 3 - 4 independent observations taken over a night.
The light curve during the observing years 2001-2002 has a
dense temporal coverage. We, therefore, divided this light curve into two different epochs to
see any variation in the $\theta_{min}$ and the amplitude.
The mean epoch of the light curves, the observed maximum ($\Delta V_{max}$)
and minimum ($\Delta V_{min}$) in the V band,  peak to peak amplitude 
($\Delta V =  \Delta V_{max} - \Delta V_{min}$), and phase of minima($\theta_{min}$)
are listed in Table \ref{ligper.tab}.
The value of $\Delta V_{max}$ was constant during each epoch indicating that the
brightness of unspotted photosphere was constant from epoch to epoch. However,
the value of $\Delta V_{min}$ was reduced by 0.14 mag from epoch 'a' to epoch 'b',
and remained constant during the epoch 'c', 'd' and 'e'.  

The RS CVn systems usually show one or two well defined minima, thereby indicating that the
rotational modulations caused by one or two prominent spots or groups of spots.
Additional spots may be present at other longitudes, or in the circumpolar regions 
but contribution to the overall rotational modulation may not be appreciable.
The phase of the light minimum ($\theta_{min}$) directly indicates the mean longitude of the
dominant groups of spots. 
A sharp minimum
was observed during the epoch 'a' (see Fig. \ref{fold.fig}a). At the same time
the amplitude of the V band light curve was found to be
0.288 mag, which was  maximum  during our observations. The
sharpness of the minimum indicate that it is  the latitudinal extent of the 
groups of spots that may be responsible rather than the  
longitudinal extent. Broad minima during the epochs 'b', 
'c' and 'd' indicate that the spots were spread over an appreciable longitudinal range
(Fig. \ref{fold.fig} b to d). 
It is interesting to see the light curve of the star HD 81032 during the epoch 'e'. 
Here a single large spot, characterized by a broad minimum during the epochs 'b', 'c' and 'd'
separated into two groups of spots. This can be easily seen by two well separated minima
(see Fig. \ref{fold.fig} e).  Significant change in $\theta_{min}$ 
(see Table \ref{ligper.tab}) is 
probably associated with a change in the spot configuration on the surface 
of the star.

 Variation in the colour of the star HD 81032 is correlated with its magnitude (see top
two panels of Fig. \ref{fold.fig} a to e)
i.e. the star becomes redder when fainter, and bluer when brighter,
supporting the starspot hypothesis.  The significance of the correlation
has been calculated by determining the linear correlation coefficient, r,
between the magnitude and the colours.
The value of r between V and (B-V), V and (V-R) and (B-V) and (V-R) was found
to be 0.20, 0.55 and 0.40, with the corresponding probability
of no correlation being 0.0253, $6.157 \times 10^{-12}$ and $2.443\times10^{-6}$, respectively.
The conventional starspot 
model assumes that the spots are cooler than the surrounding photosphere, and hence one would
expect the star to be the reddest at the light minimum, although 
flaring activity can lead an opposite effect being observed 
as seen in UX Ari (Ulv\aa s \& Henry 2003; Padmakar \& Pandey 1999; Ravindran \& Mohin 1995).

\begin{figure}
\centering
\includegraphics[height=10.8cm,width=11.8cm]{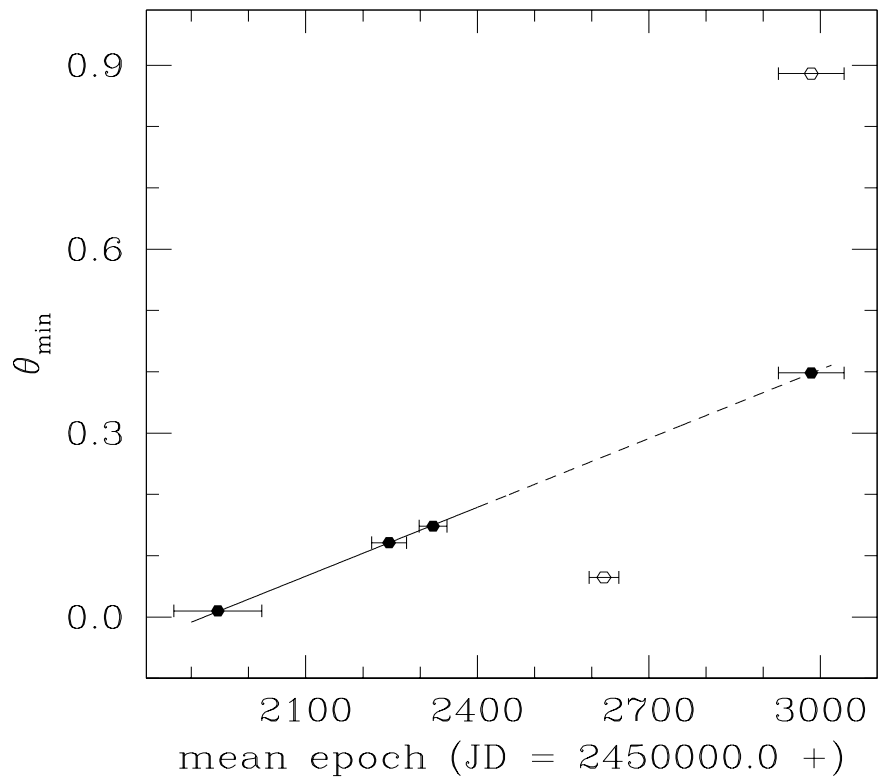}
\caption{Plot of mean epoch versus light minima. The solid line 
represents the linear least square fit of the eq. \ref{migra.eq}. The open
circles were not used in the fit. The horizontal bar represent the length of the
epoch.}
\label{phmin.fig}
\end{figure}

Figure \ref{phmin.fig} shows the plot of the mean epoch versus phase minima ($\theta_{min}$) of 
the V band light curves.  The following equation  was fitted to the data by the
method of linear least squares to determine the rate of the phase shift
(or migration period).  

\begin{equation}
\theta_{min} = \omega (t - t_{0}) + \theta_{0}\,,
\label{migra.eq}
\end{equation}

\noindent
Here $\omega = 2 \pi/P$ is the
rate of phase shift (degrees/day), P is the corresponding period in days,
t is time in days, $t_{0}$ is reference time, and $\theta_{0}$ is the
reference spot longitude in degree.
After the epoch JD 2452400 (or after the epoch 'd'), the evolution of
different spots seems to be more responsible than
the migration of single group of spots, so, we have used only first three points
of the Fig. \ref{phmin.fig} (i.e. the phase minima of epoch 'a', 'b' and 'c') 
in the linear least square fit of equation (1). The solid line in Fig. 
\ref{phmin.fig} shows the linear least square fit of the equation (1) to the 
data. The rate of phase shift for the group of spots was determined as 
$0.1348 \pm 0.002$ degree/day, which  corresponds
to a  migration period of $7.43 \pm 0.07$ years.
The smaller group of spots during the epoch 'e' appears to rotate with the 
similar angular velocity with that of a group of spots during the epoch 'a', 'b'
and 'c' (see the dotted line in Fig. \ref{phmin.fig}).
However, long-term studies of spot groups' evolution and migration in the star
HD 81032 require much more frequent sampling in the time domain.
The arrangement of 
 groups of spots in one permanent strip can be easily interpreted as 
a long lived active group rotating with the same velocity.
The open circles in the Fig. \ref{phmin.fig} are not following that strip, and may be 
due to the formation of a new group of spots  on the surface of the star.

\section{H$\alpha$ and CaII H and K emission lines}
\label{halcai.sec}
\begin{figure}
\centering
\includegraphics[height=10cm,width=12cm]{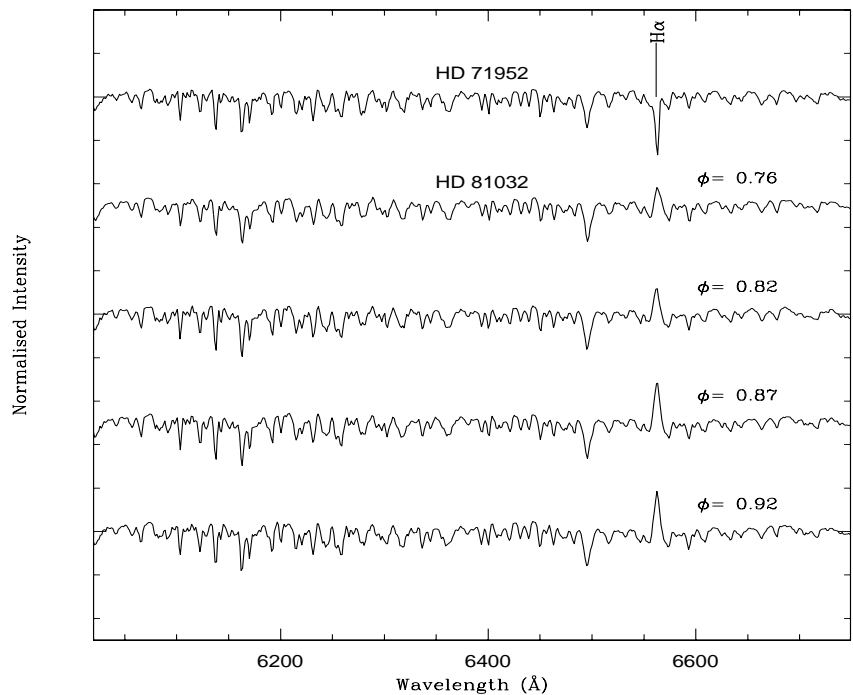}
\caption{H$\alpha$ spectra of HD 81032. HD 71952 was taken as reference spectra.}
\label{halpha.fig}
\end{figure}

\begin{figure}
\centering
\includegraphics[height=10cm,width=12cm]{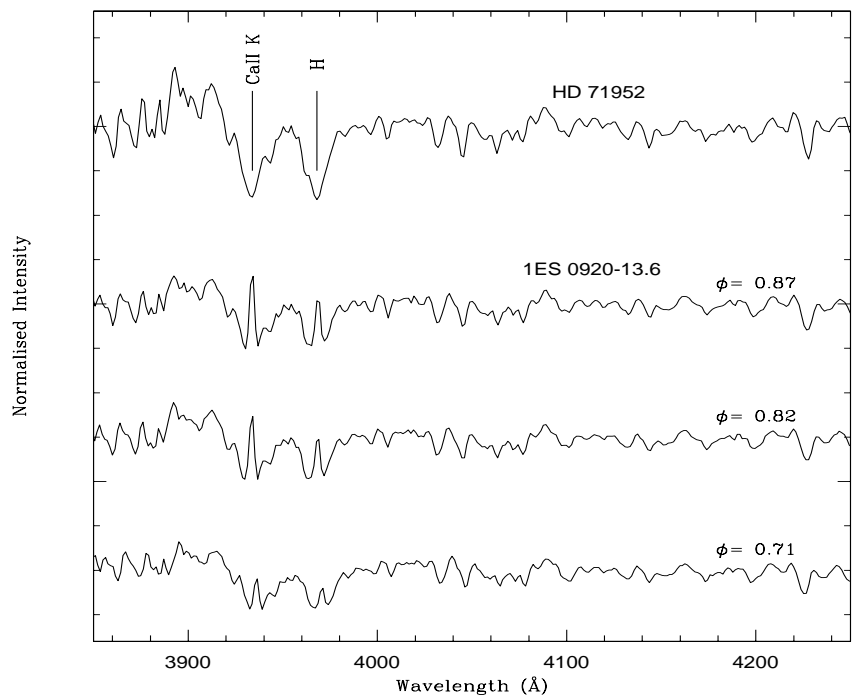}
\caption{$CaII H \& K$ spectra of HD 81032. Spectra of HD 71952 was taken as a reference.}
\label{caiihk.fig}
\end{figure}

The H$\alpha$ and CaII H \& K emission lines are important indicators of chromospheric activity.
In very active stars like II Peg,
V711 Tau, UX Ari the  H$\alpha$ emission is seen clearly above the continuum,
whereas  in less active stars only a filled in absorption line is observed.
 Figure \ref{halpha.fig} shows spectra of HD 81032 in the H$\alpha$ region.
Spectrum of the star HD 71952 is shown for comparison.
H$\alpha$ emission line can be seen clearly in the star HD 81032 (Fig. \ref{halpha.fig}).
It has been suggested that in the RS CVn systems at least one star must also show the 
intense emission in the \ H and \ K lines of CaII (Fekel et al. 1986).
As shown in Figure \ref{caiihk.fig}, CaII H and K lines are above the continuum
at all the phases observed. The measured equivalent widths (EWs) of
the H$\alpha$, CaII H and K emission feature, the corresponding JD and phase are listed
in Table \ref{ew.tab}. 

\begin{figure}
\centering
\includegraphics[height=8cm,width=11cm]{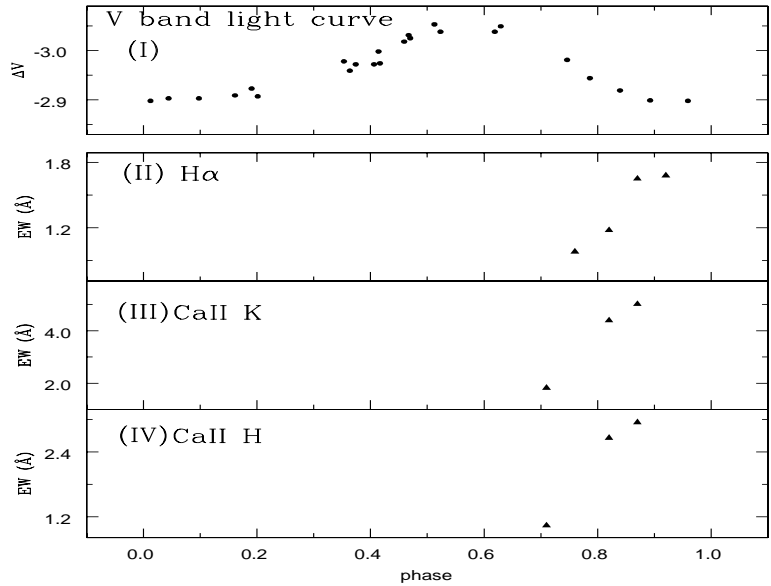}
\caption{(I) V light curve of the HD 81032 during the epoch 'd', (II)
variation of EWs of H$\alpha$ emission feature, (III) CaII K EWs  and (IV)
Ca II H EWs.  Phases are reckoned from JD 2452307.761 and the period $18^{d}.802$. }
\label{splc.fig}
\end{figure}

The photometric observations during the 
epoch 'd' were close to the spectroscopic observations. 
Figure \ref{splc.fig} shows the plot of
EWs of Ca II \ H, \ K and H$\alpha$ against phase, along with the V band light curve of the epoch 'd'.
It appears that H$\alpha$, Ca II H, and K  emission features 
are variable and anti-correlated with the photometric phase i.e. the maximum at photometric
minimum and minimum at photometric maximum. This could be due to the presence of active cool spots 
on the surface of the star.

\section{Spectral type}
\label{physical.sec}
The  value of the total galactic reddening E(B-V) in the direction of HD 81032
is estimated to be 0.04 mag from the reddening map given by Schlegel et al. (1998).
However, the  star HD 81032 suffers negligible reddening at a distance of
140 pc, and we assume E(B-V) = 0 for the remaining discussion. 
Using the
Tycho parallax (Hog 1997) and the value of $V (=8.63 ~\rm{mag})$ mag from the present
photometry the absolute magnitude $M_V$ of HD 81032 is 3.04 mag. 
 The value of $M_V$ is  consistent with
the luminosity class IV for this star. The $(B-V)$ colour
of HD 81032 derived from our photometry  is $1.02\pm0.01$ consistent with the relatively 
precision to the  value of $0.98\pm0.05$ given in the Tycho catalogue (Wright et al. 2003).
The (B-V) colour is best
matched with the spectral class K0IV, and is consistent with the spectral class 
identified by Wright et al. (2003).
Applying the bolometric correction -0.4 (Schmidt \& Kaller 1982) for K0IV star,
the bolometric magnitude of HD 81032 is 2.65 mag. 

\section{Spectral energy distribution (SED)}
\label{sed.sec}
\begin{figure}
\centering
\includegraphics[height=9cm,width=11cm]{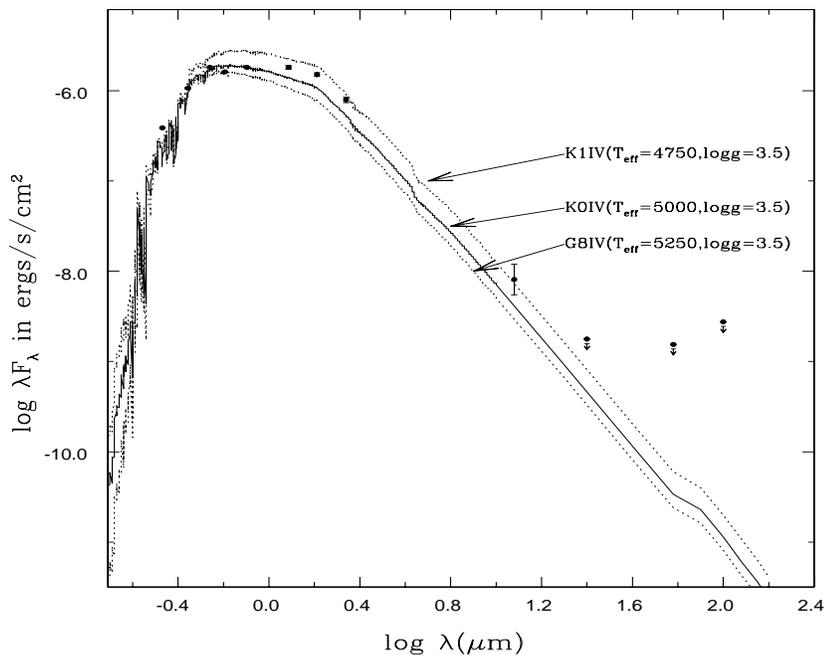}
\caption{SED of the star HD 81032 (solid dots). The solid line represents
the best match model SED from Kurucz (1993) as expected from the intrinsic properties of the star, dotted lines represent the SED of the K1IV and G8IV type star.}
\label{sed.fig}
\end{figure}

We have determined the SED of HD 81032 using broad band UBVRI (present photometry),
2MASS JHK (Cutri et al. 2003) and 12, 25, 60 and 100 $\mu$ IRAS (Moshir 1989)
fluxes. However, only upper limits are available at 25, 60 and 100 $\mu m$. The observed SED of HD 81032 
along with the synthetic SED is shown in Figure \ref{sed.fig}. The synthetic SED is expected
from the  intrinsic properties of the star (Kurucz 1993). The model SED shown 
in Fig. \ref{sed.fig} has been adjusted to coincide with the observed SED 
of the star HD 81032 at the V-band wavelength of 0.55 $\mu$.
We have  overplotted the synthetic SED for the different $T_{eff}$ and $\rm{log} g$ 
combinations.
The values of $T_{eff}$ and $\rm{log}g$ which best match the
observed SED are $5000\pm250$ K and $3.5\pm0.5$, respectively, and are 
consistent within one subclass of the  inferred K0IV spectral type of the star HD 81032 (see \S \ref{physical.sec}). 

The continuum of  HD 81032 is in good agreement with the normal values 
for the spectral type of K0IV up to JHK band but deviates in all bands
longer than $25 \mu m$ (see Fig. \ref{sed.fig}). The difference between the observed and the expected  (J-K) and (H-K) 
colours are $0.06 \pm 0.04$ and $0.005 \pm 0.03$ mag, respectively. This  indicates
that there is no significant colour excess in J, H and K band. The expected (J-K) and (H-K) colours were
taken from the spectral type of the star (Koorneef, 1983).
The 12 $\mu m$ magnitude ($m_{12}$) of the star HD 81032 was found to be $6.09 \pm 0.17$ mag using the 
relation $m_{12} = -2.5 \rm{log} f_{12} + 3.63$ (Mitrou et al. 1996).
The intrinsic $[K-12]$ colour of the star HD 81032 is determined to be $0.03\pm 0.17$ mag.
The expected $[K-12]$ colour of 0.14 mag for the K0IV type star
(Verma et al. 1987) gives the colour excess of 0.17 mag, which is 
consistent with the ideal value of $0.0\pm0.2$ mag (Mitrou et al. 1996).

\section{X-ray light curve and spectra}
\label{xspec.sec}

\begin{figure}
\centering
\includegraphics[height=14cm,width=9cm,angle=-90]{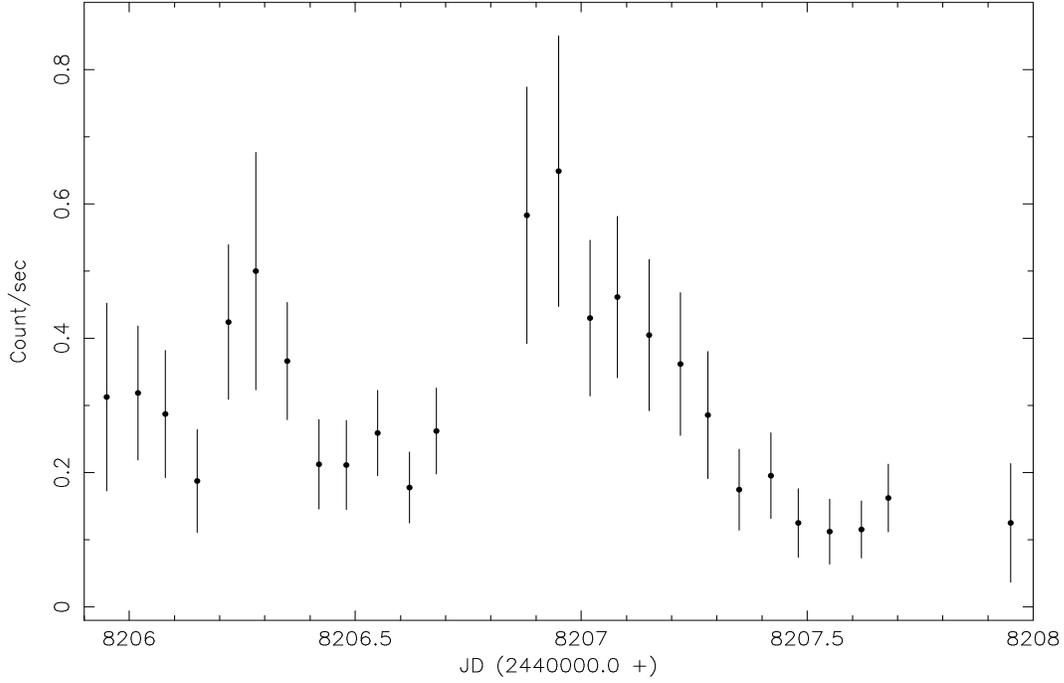}
\caption{Background subtracted RASS PSPC X-ray light curve of HD 81032.
The bin size in the light curve is 64 s.}
\label{xlc.fig}
\end{figure}

The light curve for the source and background were extracted using 
the xselect package for the PSPC 0.1 - 2.4 keV energy band which contains
all the X-ray photons. The background-subtracted X-ray light curve of 
HD 81032 is shown in Figure \ref{xlc.fig} plotted with a time bin size 
of 64 s. It appears from the light curve that a moderate flare occurred 
during the RASS observations, with a peak of about 0.6 ct s$^{-1}$ 
(compared to a pre-flare level of 0.2 - 0.3 ct s$^{-1}$) at 
approximately JD = 2448206.95 (1990/11/11 10:48:0.0) and a half-decay time of
$2.6 \times 10^{4}$ s. 

\begin{figure}
\centering
\includegraphics[height=14cm,width=9cm,angle=-90]{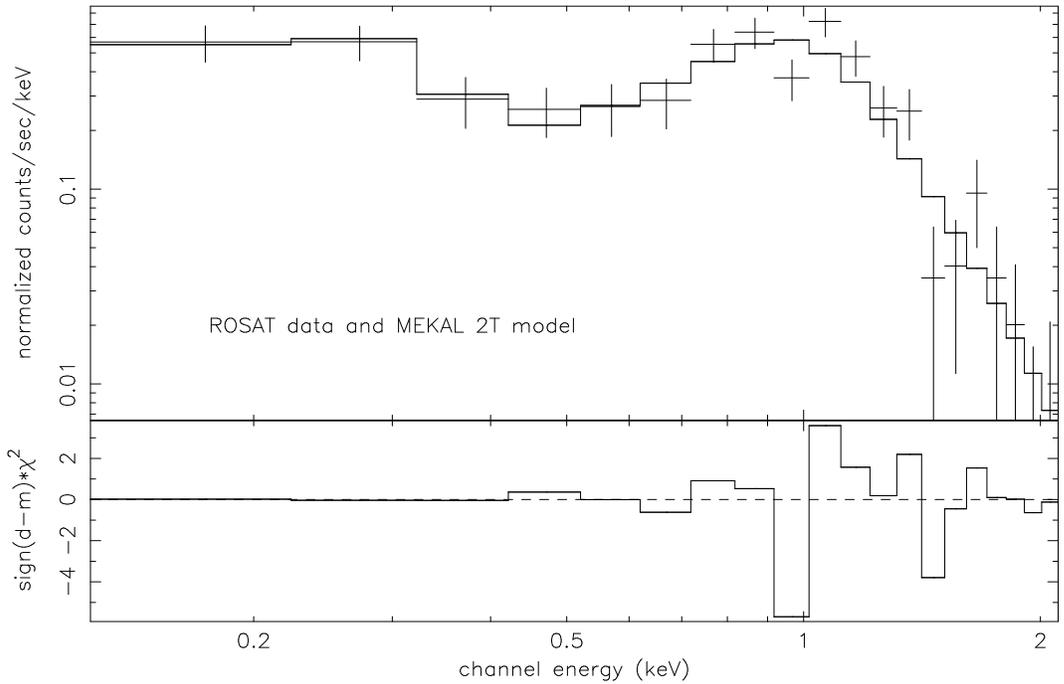}
\caption{Spectra of HD 81032 taken with the ROSAT PSPC detector (top panel),
the bottom panel shows the $\chi^{2}$ for each bin.}
\label{xspec.fig}
\end{figure}

The X-ray spectrum of HD 81032 as observed with the ROSAT PSPC 
is shown in  Figure \ref{xspec.fig}. 
Response matrices based on the available off-axis calibration of the PSPC and
using the appropriate ancillary response files were created and data were fitted using the {\it xspec}
(version 11.3.1) spectral analysis package. Spectral
models for thermal equilibrium plasma known as the Mewe-Kaastra-Liedahl or MEKAL
model (Liedahl, Ostrecheld \& Goldstein 1995; Mewe, Kaastra \& Liedahl 1995) were used.
The background-subtracted X-ray spectra were fitted with 1-temperature (1T) and  
2-temperature (2T) plasma models, 
either assuming solar photospheric abundances as given by Anders \& Grevess (1989)
or allowing the abundance of every element other than H to vary by a
common factor relative to the solar (photospheric) values. In each of the
above models the interstellar absorption was assumed to follow the absorption
cross-sections given by Morrison \& McCammon (1983), and 
the total intervening hydrogen column density $N_H$ was allowed to vary freely.

The results of different model fits are summarized in Table \ref{xdata.tab}. 
Single-temperature  MEKAL models with abundances fixed to 
the solar values gave unacceptably high values for $\chi_{\nu}^{2}$, and 
thus can be rejected. However,
single-temperature plasma models with abundances of $0.19_{-0.08}^{+0.14}$
times solar and plasma temperature of $0.84_{-0.20}^{+0.17}$  were found acceptable.
Alternatively, two-temperature MEKAL models with fixed solar abundances were 
also found to be acceptable; the  two
temperatures being  $0.2_{-0.1}^{+0.2}$ keV and $1.12_{-0.36}^{+0.52}$ keV. 
Multi-temperature variable abundance models would also
produce acceptable fits to the PSPC spectrum, but the small number of
accumulated counts does not warrant such complex models. 
The PSPC spectrum and the best-fit two-temperature plasma 
model with solar abundances are shown in Fig. \ref{xspec.fig}
along with the significance of the residuals
in terms of their $\chi^{2}$.

Based on the best fit 2T MEKAL model a source flux of $9.1_{-1.6}^{+0.9} \times 10^{-12}$ 
erg cm$^{-2}$ s$^{-1}$ was obtained. At a distance of 140 pc the
 X-ray luminosity of  HD 81032 is calculated 
to be $2.1_{-0.4}^{+0.2} \times 10^{31} \rm {erg} \rm{s}^{-1}$, similar 
to that of known subgiants RS CVn binaries (Drake et al. 1989, Singh et al. 1995,1996; 
Dempsey et al. 1997; Padmakar et al. 2000; Pandey et al. 2005).
The derived X-ray luminosity
is about a factor of 2 bigger than the earlier estimate given in \S1.
This may be due to the non-trivial column density of $1.4 \times 10^{20}$
$cm^{-2}$ inferred: objects with a high column density will typically have a
higher count rate to flux conversion factor than the standard one which
assumes a low column density.
The values of $kT_{1}$ and $kT_{2}$ of HD 81032 (see Table \ref{xdata.tab}) are consistent with 
those of other RS CVn systems studied by Dempsey et al. (1997), who found an 
average value of
 $0.18\pm 0.01$ keV and $1.37\pm 0.04$ keV for a sample of 28 RS CVn systems. 
The volume emission measures  EM1 and EM2 of  HD 81032
are also found to be  consistent with those of the other similar systems,
 where the average volume emission measures $EM_{1}$ and $EM_{2}$
are $1.2\pm0.3 \times 10^{53}$ and $5.7\pm1.2 \times 10^{53}$ cm$^{-3}$,
respectively (Dempsey et al. 1997). 

All the errors
quoted in this section are internal statistical errors, and that the
distance-dependent errors of 45\% in luminosity which were mentioned in \S
1 have not been included in, for example, the EM errors quoted in Table 3
or the errors in the X-ray luminosity given the above paragraph of this
section.

\section{Summary}
\label{con.sec}
A photometric period of $18.802 \pm 0.074$d has been discovered in the star HD 81032.
The amplitude and the phase of the photometric light curves of HD 81032
are observed to be changing from one epoch to another. The amplitude 
of variation is found to be maximum (0.29 mag) during the observing 
year 2000-2001. The change in the amplitude is mainly due to a change in
the minimum of the light curve, and this may be due to a change in the spot coverage
on the surface of the star.  A single group of spots is found to migrate,
and a migration period of $7.43\pm0.07$ years is determined from the first three
years of the data. The star is seen 
to be redder at the light minimum and we interpret this is due to the relatively
cooler temperature of the darker regions present in the visible hemisphere.
A spectral type of K0IV is determined for HD 81032.

The equivalent widths of H$\alpha$, Ca II H and  K emission lines in the spectra of HD 81032 
appear more intense close to the photometric light minimum. This 
could be attributed  to  the presence of cool spots on the surface of the stars.
The observed X-ray spectrum and the inferred coronal plasma parameters for HD 81032 are
typical of those seen in active stars such as RS CVn binaries.

All of the optical and X-ray properties found for HD 81032 are most consistent 
with it being
an evolved RS CVn binary of the long-period type. Although there are a handful
of apparently single evolved stars, e.g., FK Com, which can exhibit similar
properties, identification of HD 81032 as such a rare object seems rather improbable.
Clearly, high-resolution optical spectroscopy of HD 81032 in order to
confirm the presence of radial velocity variability and determine a spectroscopic
binary orbit would be definitive evidence for its proposed binary nature.

\section*{Acknowledgment}
We are grateful to the anonymous referee for valuable comments and suggestions.
We are thankful to the time allocation committee  for giving time
at 234-cm VBT.
This research has made use of data obtained from the High Energy Astrophysics Science
Archive Research Center (HEASARC), provided by NASA's Godard Space Flight center.
Starlink is funded by PPARC and based at the Rutherford Appleton Laboratory, which is part of Council for the Central Laboratory of the Research Councils, UK.
This publication makes use of data products from the Two Micron All Sky Survey, which is a joint project of the University of Massachusetts and the Infrared Processing and Analysis Center/California Institute of Technology, funded by the National Aeronautics and Space Administration and the National Science Foundation.

\newpage 

\newpage
\begin{table}
\centering
\caption{Parameters determined from the light curves of the star HD 81032.}
\label{ligper.tab}
\begin{tabular}{lccccc}\hline
Mean epoch&Amplitude&$\Delta V_{min}$&$\Delta V_{max}$&\multicolumn{2}{c}{phase
minima}\\
\cline{5-6}
245000.0+&&&&I&II\\
\hline
&&&&&\\
(a) 1946.3350& 0.288& -2.739& -3.027 &  0.01  &  \\
&&&&&\\
(b) 2245.8599& 0.160& -2.893& -3.053 &  0.12 &\\
&&&&&\\
(c) 2322.6527& 0.156& -2.896& -3.052 &  0.15 &\\
&&&&&\\
(d) 2621.4426& 0.152& -2.890& -3.042 &  0.07 &\\
&&&&&\\
(e) 2983.8913& 0.175& -2.891& -3.066 &  0.89& 0.40\\
&&&&&\\
\hline
\end{tabular}
\end{table}

\begin{table}
\centering
\caption{Equivalent widths (EWs) of $CaII H\&K$ and 
H$\alpha$ emission lines}
\label{ew.tab}
\begin{tabular}{lcccc}\hline
JD&Phase&\multicolumn{3}{c}{EWs ($\AA$)}\\
\cline{3-5}
245000+&&$CaII K$ &$H$ &H$\alpha$ \\
\hline
&&&&\\
2659.626&0.71&1.828&1.037&   -   \\
&&&&\\
2660.513&0.76&-    & -   &$0.980$\\
&&&&\\
2661.520&0.82&4.393&2.661&$1.178$\\
&&&&\\
2662.518&0.87&5.012&2.949&$1.651$\\
&&&&\\
2663.536&0.92&    -&    -&$1.680$\\
&&&&\\
\hline
\end{tabular}
\end{table}

\newpage
\begin{table}
\centering
\caption{Results of X-ray spectral analysis}
\label{xdata.tab}
\begin{tabular}{ccccccccc}\hline
&&&&&&&&\\
Model   &Abundances$^a$ &$N_H$             &$kT_1$&$EM_1$&$kT_2$&$EM_2$&$\chi_{\nu}^{2}$&Degrees of \\
        &               &$10^{20} cm^{-2}$ &(keV) &$10^{52} cm^{-3}$&(keV)&$10^{53} cm^{-3}$&&freedom\\
&&&&&&&&\\
\hline
&&&&&&&&\\
MEKAL 1T&1.0 (fixed)          &$0.21_{-0.16}^{0.22}$&$0.85_{-0.10}^{+0.29}$&$5.2_{-2.5}^{+2.2}$ &&&1.87&17\\
&&&&&&&&\\
MEKAL 1T&$0.19_{-0.08}^{0.14}$&$1.5_{-1.0}^{+1.0}$  &$0.84_{-0.17}^{+0.20}$&$18.9_{-9.9}^{+14.7}$&&&1.53&16\\
&&&&&&&&\\
MEKAL 2T&1.0 (fixed)          &$1.4_{-1.1}^{+5.4}$  &$0.2_{-0.1}^{+0.2}$&$2.9_{-1.5}^{+5.7}$ &$1.12_{-0.36}^{+0.52}$&
$1.8_{-1.6}^{+1.1}$&1.4&15\\
&&&&&&&&\\
\hline
\end{tabular}
$^a$ Common value of abundances for all the elements respect to the solar photosphere values;

 Errors are with 90\% confidence for single parameter based on $\chi_{min}^{2} + 2.71$; distance = 140 pc.
\end{table}
%-----------------------------------------------------

\end{document}